\documentclass[pdflatex,sn-mathphys-num]{sn-jnl}

\usepackage{graphicx}%
\usepackage{multirow}%
\usepackage{amsmath,amssymb,amsfonts}%
\usepackage{amsthm}%
\usepackage{mathrsfs}%
\usepackage[title]{appendix}%
\usepackage{xcolor}%
\usepackage{textcomp}%
\usepackage{manyfoot}%
\usepackage{booktabs}%
\usepackage{algorithm}%
\usepackage{algorithmicx}%
\usepackage{algpseudocode}%
\usepackage{listings}%
\usepackage{braket}
\usepackage{amsfonts}
\usepackage{color}
\usepackage{amssymb}
\usepackage{mathrsfs}
\usepackage{graphicx}
\usepackage{lmodern}
\usepackage{lineno}
\usepackage{hyperref}
\usepackage[normalem]{ulem}
\usepackage{soul}
\usepackage{amsmath,amssymb,amsfonts,bm}
\usepackage{mathtools}

\newcommand{\beq}{\begin{eqnarray}}
\newcommand{\eeq}{\end{eqnarray}}

\newcommand{\End}{\mathrm{End}}
\newcommand{\tr}{\mathrm{tr}}

\theoremstyle{thmstyleone}%
\theoremstyle{thmstyletwo}%

\theoremstyle{thmstylethree}%

\begin{document}

\title[The Geometry of Quantum Teleportation in Phase Space]{The Geometry of Quantum Teleportation in Phase Space}


\author*[1]{\fnm{Jasel} \sur{Berra-Montiel}}\email{jasel.berra@uaslp.mx}
\equalcont{These authors contributed equally to this work.}

\author[1]{\fnm{Alberto} \sur{Molgado}}\email{alberto.molgado@uaslp.mx}
\equalcont{These authors contributed equally to this work.}

\author[1]{\fnm{Mar} \sur{S\'anchez-C\'ordova}}\email{maria.cordova@uaslp.mx}
\equalcont{These authors contributed equally to this work.}

\affil*[1]{\orgdiv{Facultad de Ciencias}, \orgname{Universidad Autónoma de San Luis Potosí}, \orgaddress{\street{Campus Pedregal, Av. Parque Chapultepec 1610, Col. Privadas del Pedregal}, \city{San Luis Potosí}, \postcode{78217}, \state{SLP}, \country{Mexico}}}

\abstract{
We formulate the standard quantum teleportation protocol within the framework of deformation quantization, employing the Stratonovich--Weyl (SW) correspondence and the associated star product on the coadjoint orbits of $SU(2)$. Each qubit is represented by a spherical phase space $S^2$ endowed with the Kirillov--Kostant--Souriau (KKS) symplectic structure. Within this setting, we explicitly construct the phase space representations of the entangled Bell resource, the controlled-NOT and Hadamard operators, the projective measurements, and the conditional Pauli corrections. We demonstrate how the quantum information initially encoded in the Wigner function for the input qubit is redistributed through bipartite and tripartite phase space correlations.  Upon consideration of a conditional unitary transformation, the measurement outcome produces a Wigner function on the receiver's coadjoint orbit that is related to the input by an outcome-dependent rotation, which can subsequently be inverted through the implementation of an appropriate Pauli correction.
The covariance of the SW correspondence provides a geometric interpretation of this process, in which quantum information is redistributed and reconstructed without transferring the underlying symplectic geometry. Our formulation can also be extended to study quantum teleportation in more general phase space geometries equipped with suitable star products.}

\keywords{Deformation quantization, star product, quantum teleportation, quantum information}

\maketitle

\section{Introduction}
\label{Introduction}

Quantum teleportation is one of the fundamental protocols of quantum information theory. It provides a fundamental manifestation of the role played by entanglement and classical communication in the transmission of quantum information. In the standard protocol, an unknown quantum state initially associated with one subsystem is reconstructed in a spatially separated subsystem by combining a shared entangled resource, local unitary operations, measurement, and classical communication \cite{EPR,Mann,Zeilinger,Boschi,Nielsen}. Remarkably, the information characterizing the unknown state is not communicated through a classical measurement outcome. Rather, the entangled resource and the correlations generated during the protocol allow the state to be reconstructed at the receiver after an appropriate conditional unitary transformation. 

The conventional formulation of quantum teleportation is expressed entirely in a Hilbert space, where quantum states are represented by vectors or density operators and the protocol is described through unitary gates, projective measurements, and partial traces \cite{Nielsen}. An alternative description is provided by phase space quantum mechanics, in which quantum states and operators are represented by functions on an underlying classical phase space, while operator multiplication is replaced by a noncommutative star product \cite{Wigner,Bayen,Curtright,Prata,Generating}. Such a formulation is particularly compelling for finite-dimensional systems because their phase spaces possess a natural geometrical structure.

In the standard protocol, Alice possesses a qubit $A$, given by $\ket{\psi}_{A}\in\mathbb{C}^{2}$, in an unknown state
\begin{equation}
\ket{\psi}_A=\alpha|0\rangle_A+\beta|1\rangle_A,
\end{equation}
with $\alpha,\beta\in\mathbb{C}$, together with a qubit $B$, which forms a maximally entangled pair with Bob's qubit $C$,
\begin{equation}
\ket{\Phi^+}_{BC}
=\frac{1}{\sqrt2}\left(\ket{00}_{BC}+\ket{11}_{BC}\right)\in \mathbb{C}^{2}\otimes\mathbb{C}^{2}\simeq \mathbb{C}^{4}.
\end{equation}
Starting from the composite state, $\ket{\Psi_{0}}_{ABC}=\ket{\psi}_A\otimes\ket{\Phi^+}_{BC}\in\mathbb{C}^{8}$, Alice couples the unknown state to the entangled resource by applying a controlled-NOT operation with the qubit $A$ as the control and the qubit $B$ as the target, followed by a Hadamard transformation on $A$. These operations redistribute the information initially encoded locally in qubit $A$ into correlations among the three qubits. Alice then measures the qubits $A$ and $B$ in the computational basis and communicates the resulting two classical bits to Bob. The measurement outcome determines which of four Pauli related conditional states is obtained by qubit $C$. Once this classical information is received, Bob applies the corresponding Pauli correction, recovering the original state $\ket{\psi}$.

Within the deformation quantization framework, the corresponding phase space for a qubit can be identified with an $SU(2)$ coadjoint orbit, which is geometrically a two-sphere $S^2$
endowed with the Kirillov--Kostant--Souriau (KKS) symplectic structure \cite{Kostant,Kirillov,Kirillov2,BerraSQ}.
The Stratonovich--Weyl (SW) correspondence establishes an invertible mapping between operators acting on the qubit Hilbert space and functions defined on the spherical phase space \cite{Stratonovich,Tilma}. In particular, a density operator is mapped to a continuous Stratonovich--Weyl function, which plays the role of the Wigner function for the qubit. Since the SW correspondence is invertible, the Wigner function contains the complete information required to reconstruct the corresponding quantum state. Thus, the information conventionally encoded in the density operator is equivalently represented by a continuous function defined over the coadjoint orbit. In this sense, the Wigner function is not merely a phase space visualization of the quantum state, but a faithful and complete representation of it \cite{Varilly,Rundle}.

It is worth noting that alternative phase space formulations of quantum teleportation have previously been considered in continuous and discrete settings \cite{Janszky,Paz}. In Ref.~\cite{Janszky}, a unified Wigner function description of the quantum teleportation protocol was formulated for systems of arbitrary dimension, building on the phase space formalism introduced in Ref.~\cite{Wooters}, with particular emphasis on finite-dimensional Hilbert spaces and their continuous limit. Within this framework, the different stages of the teleportation protocol are represented in terms of transformations of Wigner functions, providing a phase space description in which imperfections in the protocol can also be incorporated. Subsequently, in Ref.~\cite{Paz}, quantum teleportation was formulated in terms of a discrete Wigner function for finite-dimensional quantum systems. This construction was particularly suited to composite systems and provided a phase space framework in which correlations and entanglement between subsystems can be analyzed. Moreover, the discrete phase space formulation was connected with the direct tomographic measurement of the corresponding Wigner function \cite{MankoSpin,MankoTomography}.

These approaches demonstrate that quantum teleportation can be expressed in terms of Wigner functions and phase space correlations. The deformation quantization formulation developed here, however, provides an additional geometrical structure that makes it possible to examine the protocol beyond its phase space representation. The phase space of each qubit is identified with an $SU(2)$ coadjoint orbit, naturally endowed with the KKS symplectic structure, while the noncommutative composition of quantum observables is encoded by its corresponding star product \cite{BerraSQ}. Within this framework, unitary transformations acquire a direct geometrical meaning as rotations of state-dependent functions over the orbit, and correlations between qubits are represented as nonseparable geometric structures over products of coadjoint orbits. This provides a natural setting for investigating how the quantum information initially encoded locally in the unknown state is geometrically redistributed throughout the different stages of the teleportation protocol. As we show, the geometrical information carried by the initial Wigner function is progressively transferred from a local phase space configuration into multipartite correlations and, after measurement and classical communication, is reconstructed as a local configuration on the receiver's coadjoint orbit. In this way, this formulation reveals a geometrical information flow underlying quantum teleportation that is not manifest in its conventional Hilbert-space description.

The paper is organized as follows. In Sec.~\ref{sec:SW_qubits}, we review the Stratonovich--Weyl correspondence for qubits, emphasizing the geometrical structure of $SU(2)$ coadjoint orbits and the associated star product, and extend this formulation to three qubit systems. In Sec.~\ref{sec:teleportation_PS}, we formulate the quantum teleportation protocol in phase space and investigate how quantum information is redistributed and reconstructed throughout the protocol. In Sec.~\ref{sec:geometrical_interpretation}, we discuss the geometric interpretation of these results and possible extensions to other phase space geometries and star products. Finally, we present our conclusions in the last section.

\section{Phase space of qubits and the Stratonovich--Weyl correspondence}
\label{sec:SW_qubits}

We begin this section by briefly reviewing the geometrical construction of the phasespace associated with a single qubit and its Stratonovich--Weyl (SW) representation. The construction follows the coadjoint orbit formulation
developed in Ref.~\cite{BerraSQ} and the references cited therein, where further details can be found.

For a qubit, the relevant symmetry group is $SU(2)$, with Lie algebra
$\mathfrak{su}(2)$ generated by
\begin{equation}
    J_i=\frac{1}{2}\sigma_i,
    \qquad i=1,2,3,
\end{equation}
where $\sigma_i$ denote the Pauli matrices, which satisfy
\begin{equation}
    [J_i,J_j]=i\epsilon_{ijk}J_k.
\end{equation}
The Killing form $\kappa(X,Y)=-2\operatorname{Tr}(XY)$, with $X,Y\in\mathfrak{su}(2)$, defines a nondegenerate invariant bilinear form on $\mathfrak{su}(2)$, establishing a natural identification $\mathfrak{su}(2)\simeq\mathfrak{su}(2)^*$, where $\mathfrak{su}(2)^*$ denotes the dual space of the Lie algebra. Under this identification, the coadjoint action of $SU(2)$ on $\mathfrak{su}(2)^*$ corresponds to the adjoint action on $\mathfrak{su}(2)$. Since the adjoint representation of $SU(2)$ is equivalent to the rotational action of $SO(3)$, the norm of an element of $\mathfrak{su}(2)^*$ is preserved. Consequently, the nontrivial coadjoint orbit on any nonzero element $\xi\in\mathfrak{su}(2)^{*}$ is a two-dimensional sphere,
\begin{equation}
    \mathcal{O}_{\xi}
    =
    \left\{
    \operatorname{Ad}^{*}_{g}\xi
    \,;\,
    g\in SU(2)
    \right\}
    \simeq S^2.
\end{equation}
Each coadjoint orbit carries a canonical symplectic structure, known as the Kirillov--Kostant--Souriau (KKS) form \cite{Kostant,Kirillov,Kirillov2}, given by
\begin{equation}
    \omega_{\xi}
    \left(
    \operatorname{ad}^{*}_{X}\xi,
    \operatorname{ad}^{*}_{Y}\xi
    \right)
    =
    \langle \xi,[X,Y]\rangle,
\end{equation}
where $X,Y\in\mathfrak{su}(2)$, $\xi\in\mathfrak{su}(2)$, and $\langle \cdot,\cdot\rangle$ denotes the natural dual pairing between the Lie algebra $\mathfrak{su}(2)$ and its dual $\mathfrak{su}(2)^{*}$. Thus, the sphere is not merely introduced as a geometrical representation of Bloch vectors; rather it constitutes the natural symplectic phase space of a qubit. In spherical coordinates, the KKS symplectic form coincides, up to the chosen normalization, with the standard area form on the unit sphere,
\begin{equation}
    \omega_{\mathrm{KKS}} = \sin\theta\,d\theta\wedge d\phi.
    \label{KKS}
\end{equation}
Accordingly, each point on the coadjoint orbit can be parametrized by the unit vector $\mathbf{n}\in S^2$,
\begin{equation}
    \mathbf{n}
    =
    \left(
    \sin\theta\cos\phi,
    \sin\theta\sin\phi,
    \cos\theta
    \right),
\end{equation}
with $0\leq\theta\leq\pi$ and $0\leq\phi\leq 2\pi$.

The quantization of coadjoint orbits is implemented through the Stratonovich--Weyl (SW) correspondence, which establishes a one-to-one relation between operators acting on the qubit Hilbert space and functions defined on the spherical symplectic phase space \cite{Stratonovich}. More precisely, the correspondence is defined through a family of operator-valued kernels $\hat{\Delta}(\mathbf{n})$ , with $\mathbf{n}\in S^2$, satisfying a set of properties that ensure a consistent phase space representation of the quantum theory.

For an operator $\hat{A}$ acting on $\mathcal{H}\simeq\mathbb{C}^2$, its SW symbol is defined as
\begin{equation}
    W_{\hat{A}}(\mathbf{n})=\tr \left[
        \hat{A}\hat{\Delta}(\mathbf{n})
    \right].
    \label{SW}
\end{equation}
The SW correspondence satisfies the following physically motivated postulates:

\begin{enumerate}
\item \textit{Linearity and invertibility.}
The map $\hat{A}\mapsto W_{\hat{A}}(\mathbf{n})$ is linear and one-to-one. The inverse correspondence is given by
\begin{equation}
    \hat{A}=\int_{S^2}W_{\hat{A}}(\mathbf{n})\hat{\Delta}(\mathbf{n})\,d\mu(\mathbf{n}),
    \label{eq:SW_inverse}
\end{equation}
where $d\mu(\mathbf{n})$ denotes the invariant measure on the
coadjoint orbit. Consequently, the operator and phase space
descriptions contain the same information.

\item \textit{Reality.}
The SW symbol associated with any Hermitian operator, $\hat{A}=\hat{A}^{\dagger}$, is assumed to be real, i.e., $W_{\hat{A}}(\mathbf{n})\in\mathbb{R}$. This requirement is ensured by the Hermiticity of the SW kernel. In particular, the SW symbol of a density operator defines a real quasiprobability distribution on the sphere.

\item \textit{Standardization.}
The phase space integral of the SW symbol reproduces the trace of
the corresponding operator,
\begin{equation}
    \int_{S^2}
    W_{\hat{A}}(\mathbf{n})\,
    d\mu(\mathbf{n})
    =
    \tr(\hat{A}).
    \label{eq:SW_standardization}
\end{equation}
This is equivalent to requiring that the kernel satisfies
\begin{equation}
\int_{S^2}\hat{\Delta}(\mathbf{n})\, d\mu(\mathbf{n})= I.
\end{equation}
where $I$ denotes the identity operator. For a normalized density operator, $\tr\hat{\rho}=1$, this
implies
\begin{equation}
    \int_{S^2}W_{\hat{\rho}}(\mathbf{n})\,d\mu(\mathbf{n})=1.
\end{equation}

\item \textit{Traciality.}
The Hilbert--Schmidt product between two operators is reproduced by the corresponding phase space functions,
\begin{equation}
    \operatorname{Tr}(\hat{A}\hat{B})
    =\int_{S^2}W_{\hat{A}}(\mathbf{n})W_{\hat{B}}(\mathbf{n})\,
    d\mu(\mathbf{n}).
    \label{eq:SW_traciality}
\end{equation}
This property allows to calculate operator traces and expectation values entirely in terms of phase space quantities.

\item \textit{Covariance.}
The SW correspondence is compatible with the action of the symmetry group. For $g\in SU(2)$, the kernel satisfies
\begin{equation}
    \hat{\Delta}(g\cdot\mathbf{n})
    =
    \hat{U}(g)\hat{\Delta}(\mathbf{n})\hat{U}^{\dagger}(g),
    \label{eq:SW_covariance_kernel}
\end{equation}
where $\hat{U}(g)$ corresponds to the unitary representation of $SU(2)$ acting on the qubit Hilbert space. Consequently,
\begin{equation}
    W_{\hat{U}(g)\hat{A}\hat{U}^{\dagger}(g)}(\mathbf{n})
    =W_{\hat{A}}(g^{-1}\cdot\mathbf{n}).
    \label{eq:SW_covariance}
\end{equation}
Since the adjoint action of $SU(2)$ induces rotations on the
coadjoint orbit, the action $g^{-1}\cdot\mathbf{n}$ can equivalently be expressed as $R_g^{-1}\mathbf{n}$, where $R_{g}\in SO(3)$, is the rotation associated with $g\in SU(2)$. The covariance property therefore takes the form
\begin{equation}
    W_{\hat{U}(g)\hat{A}\hat{U}^{\dagger}(g)}(\mathbf{n})=W_{\hat{A}}
    \left(R_g^{-1}\mathbf{n}\right).
    \label{eq:SW_rotation_covariance}
\end{equation}

\end{enumerate}
These properties will play a central role in the phase space formulation of quantum teleportation. In particular, invertibility ensures that reconstructing the phase space representation of the initial state on Bob's phase space is equivalent to reconstructing the original quantum state. Covariance, on the other hand, provides a geometric interpretation of the conditional unitary operations, since, conjugation by a unitary operator is mapped, through the SW correspondence, to a rotation of the corresponding phase space symbol on the coadjoint orbit.

For the spin-$1/2$ representation, the SW kernel is given by \cite{BerraSQ,Tilma},  
\begin{equation}
    \hat{\Delta}(\mathbf{n})=\frac{1}{2}\left(I+\sqrt{3}\,\mathbf{n}\cdot\boldsymbol{\sigma}\right),
    \label{SWkernel}
\end{equation}
where $\mathbf{n}$ is the unit vector on the sphere given by $\mathbf{n}=(\sin\theta\cos\phi,\sin\theta\sin\phi,\cos\theta)^{T}$, and $\boldsymbol{\sigma}=(\sigma_x,\sigma_y,\sigma_z)^{T}\equiv(\sigma_1,\sigma_2,\sigma_3)^{T}$ denotes the vector of Pauli matrices. For a linear operator $\hat{A}\in\End(\mathbb{C}^{2})$ acting on the qubit Hilbert space $\mathcal{H}\simeq\mathbb{C}^{2}$, the corresponding SW symbol is defined by
\begin{equation}
    W_{\hat{A}}(\mathbf{n})=\tr\left[\hat{A}\hat{\Delta}(\mathbf{n})
    \right].
    \label{SW_symbol}
\end{equation}
The SW correspondence is invertible. To make this explicit, we introduce the normalized invariant measure on the sphere,
\begin{equation}
d\mu(\mathbf{n})=\frac{1}{2\pi}\sin\theta d\theta d\phi,
\end{equation}
in terms of which the inverse SW transformation is given by
\begin{equation}
\hat{A}=\int_{S^2}W_{\hat{A}}(\mathbf{n})\hat{\Delta}(\mathbf{n})d\mu(\mathbf{n}).
\label{SWinverse}
\end{equation}
Thus, the SW symbol $W_{\hat A}(\mathbf n)$ contains the complete information encoded in the operator $\hat A$, so that, no information is lost in passing from the operator formulation to the phase space representation.

Since any linear operator acting on $\mathbb{C}^{2}$ can be expressed in the Pauli basis as $\hat{A}=a_0 I+\mathbf{a}\cdot\boldsymbol{\sigma}$, where $a_{0}\in\mathbb{R}$ and $\mathbf{a}\in\mathbb{R}^{3}$, the correspondign SW symbol takes the particularly simple form
\begin{equation}
    W_{\hat{A}}(\mathbf{n})=a_0+\sqrt{3}\mathbf{a}\cdot\mathbf{n}.
    \label{general_SW_symbol}
\end{equation}
In particular, a density operator $\hat{\rho}=\frac{1}{2}\left(I+\mathbf{r}\cdot\boldsymbol{\sigma}\right)$
where $\mathbf{r}$ is the corresponding Bloch vector. The SW symbol associated with $\hat{\rho}$ defines its Wigner function on the phase space $S^2$, and is given by
\begin{equation}
    W_{\hat{\rho}}(\mathbf{n})=\frac{1}{2}\left(1+\sqrt{3}\,\mathbf{r}\cdot\mathbf{n}\right).
    \label{eq:qubit_Wigner}
\end{equation}
Consequently, the complete information specifying the qubit state can be equivalently represented by a continuous function defined on the coadjoint orbit $S^2$.

Through the SW correspondence~(\ref{SW_symbol}), the phase space representation of the operator product $\hat{A}\hat{B}$, denoted by $W_{\hat{A}\hat{B}}(\mathbf{n})$, induces a noncommutative product between the corresponding SW symbols \cite{BerraSQ}. The resulting star product is given by
\begin{eqnarray}
    W_{\hat{A}\hat{B}}&=& W_{\hat{A}}\star W_{\hat{B}} \nonumber\\
    &=& a_0b_0+\mathbf{a}\cdot\mathbf{b}+
    \sqrt{3}\left(a_0\mathbf{b}+b_0\mathbf{a}+i\,\mathbf{a}\times\mathbf{b}\right)\cdot\mathbf{n}.
    \label{starQ}
\end{eqnarray}
This product can also be interpreted as a realization of the algebra of complexified quaternions \cite{Girard,Leo}.
In particular, for the generators  of $\mathfrak{su}(2)$, using $W_{\sigma_{i}}(\mathbf{n})=\sqrt{3}n_{i}$, the star product satisfies
\begin{equation}
    W_{\sigma_i}\star W_{\sigma_j}=\delta_{ij}+i\epsilon_{ijk}W_{\sigma_k},
    \label{eq:Pauli_star}
\end{equation}
which reproduces, at the level of SW symbols, the multiplication rule of the Pauli matrices. Thus, the SW star product yields an exact phase space realization of the qubit operator algebra.

We now extend the previous construction to a composite system of three qubits involved in the teleportation protocol. Let the individual subsystems be denoted by $A$, $B$, and $C$, with associated Hilbert spaces $\mathcal{H}^A$, $\mathcal{H}^B$, and $\mathcal{H}^C$, respectively. The Hilbert space of the composite system is therefore
\begin{equation}
    \mathcal{H}^{ABC}
    =
    \mathcal{H}^A\otimes
    \mathcal{H}^B\otimes
    \mathcal{H}^C
    \simeq
    \left(\mathbb{C}^{2}\right)^{\otimes 3}.
\end{equation}
For a multipartite system in which the individual qubits are regarded as distinct subsystems, it is natural to associate an independent $SU(2)$ coadjoint orbit with each of them. The phase space of the composite system is then constructed as the Cartesian product of the corresponding local phase spaces,
\begin{equation}
    \Gamma^{ABC}
    =
    (S^2)^A
    \times
    (S^2)^B
    \times
    (S^2)^C,
    \label{eq:three_qubit_phase_space}
\end{equation}
with points denoted by $ (\mathbf{n}^A,\mathbf{n}^B,\mathbf{n}^C)\in (S^2)^A\times(S^2)^B\times(S^2)^C$.
Each factor $(S^2)^j$, $j=A,B,C$, represents the coadjoint-orbit phase space associated with the corresponding qubit. This product-space description preserves the subsystem structure of the composite quantum system and provides a natural framework for representing both local observables and correlations among the individual qubits. Nevertheless, as discussed in Refs.~\cite{BerraSQ,Tilma}, an $n$-qubit system may alternatively be regarded as a $N$-level system, with $N=2^n$, whose full symmetry group is $SU(N)=SU(2^n)$. In this setting, one may associate the system with higher-dimensional coadjoint orbits of $SU(2^n)$, which, depending on the isotropy subgroup, are diffeomorphic to flag manifolds~\cite{Flag}. As coadjoint orbits, these manifolds are naturally endowed with the KKS symplectic form and, together with compatible complex and Riemannian structures, acquire the structure of K\"ahler manifolds.

For the product phase space description, the corresponding three-qubit SW kernel is then given by 
\begin{equation}
    \hat{\Delta}^{ABC}
    (\mathbf{n}^A,\mathbf{n}^B,\mathbf{n}^C)=
    \hat{\Delta}^A(\mathbf{n}^A)\otimes\hat{\Delta}^B(\mathbf{n}^B)
    \otimes\hat{\Delta}^C(\mathbf{n}^C).
\label{eq:three_qubit_kernel}
\end{equation}
where $\hat{\Delta}^{j}(\mathbf{n}^{j})$, with $j=A,B,C$, denotes the single qubit SW kernel as defined in Eq. (\ref{SWkernel}). Accordingly, the SW correspondence for the composite system is obtained from the tensor product of the local single qubit kernels. Thus, for an arbitrary linear operator
$\hat{M}^{ABC}\in\End(\mathbb{C}^{2})^{\otimes3})$ acting on $\mathcal{H}^{ABC}\simeq(\mathbb{C}^{2})^{\otimes3}$, its SW symbol is a function on the product phase space $\Gamma^{ABC}$, defined by
\begin{equation}
W_{\hat{M}^{ABC}}
(\mathbf{n}^A,\mathbf{n}^B,\mathbf{n}^C)
=
\tr\Big[\hat{M}^{ABC}
\hat{\Delta}^{ABC}(\mathbf{n}^{A},\mathbf{n}^B,\mathbf{n}^C)\Big],
\label{eq:SW3q}
\end{equation}
where the trace is taken over the full Hilbert space
$\mathcal{H}^{ABC}$.

Since the single qubit SW correspondence is invertible (\ref{eq:SW_inverse}), its tensor product extension is also invertible. Consequently, the operator
$\hat{M}^{ABC}$ can be reconstructed completely from its phase space symbol according to
\begin{equation}
\hat{M}^{ABC}
=
\int_{\Gamma^{ABC}}
W_{\hat{M}^{ABC}}
(\mathbf{n}^A,\mathbf{n}^B,\mathbf{n}^C)
\hat{\Delta}^{ABC}(\mathbf{n}^A,\mathbf{n}^B,\mathbf{n}^C)
d\mu^{ABC},
\label{}
\end{equation}
where $d\mu^{ABC}$ denotes the invariant product measure $d\mu^{A}d\mu^{B}d\mu^{C}$ on the three $SU(2)$ coadjoint
orbits. Hence, the SW function
$W_{\hat{M}^{ABC}}(\mathbf{n}^A,\mathbf{n}^B,\mathbf{n}^C)$ provides a complete phase space representation of the three qubit operator while retaining the tensor-product structure of the individual subsystems.

The product space representation also provides a convenient way of identifying the local and correlation contributions of a multipartite quantum state. Indeed, an arbitrary three qubit density operator can be expanded in the Pauli basis \cite{Nielsen},
\begin{equation}
    \hat{\rho}^{ABC}=\frac{1}{8}\sum_{\alpha,\beta,\gamma=0}^{3}T_{\alpha\beta\gamma}\,\sigma_{\alpha}^{A}\otimes
    \sigma_{\beta}^{B}\otimes\sigma_{\gamma}^{C},
    \label{eq:three_qubit_Pauli}
\end{equation}
where $\sigma_0=I$ and the real coefficients
$T_{\alpha\beta\gamma}$ are given by
\begin{equation}
    T_{\alpha\beta\gamma}=\tr\left[\rho^{ABC}\left(\sigma_{\alpha}^{A}
    \otimes\sigma_{\beta}^{B}\otimes\sigma_{\gamma}^{C}\right)\right].
    \label{eq:three_qubit_tensor}
\end{equation}
With this normalization $T_{000}=1$, while the remaining components encode the local Bloch vectors and the different  correlations among the three subsystems. By linearity of the SW correspondence (\ref{SWinverse}) and the tensor product structure of the kernel (\ref{eq:SW3q}), the phase space representation of the density operator (\ref{eq:three_qubit_Pauli}) takes the form
\begin{eqnarray}
    W_{\hat{\rho}^{ABC}}
    (\mathbf{n}^A,\mathbf{n}^B,\mathbf{n}^C)
    =
    \frac{1}{8}
    \sum_{\alpha,\beta,\gamma=0}^{3}
    T_{\alpha\beta\gamma}\,
    W_{\sigma_\alpha^A}(\mathbf{n}^A)
    W_{\sigma_\beta^B}(\mathbf{n}^B)
    W_{\sigma_\gamma^C}(\mathbf{n}^C).
\label{eq:three_qubit_Wigner}
\end{eqnarray}
Consequently, the tensor coefficients
$T_{\alpha\beta\gamma}$ acquire a direct representation as functions on the product phase space. Components with only one nonzero Pauli index determine local contributions on the corresponding sphere, components with two nonzero Pauli indices describe bipartite correlations between two phase spaces, and components with three nonzero indices describe genuine three-body contributions on the full product space.

For instance, using the notation $\sigma_i^A\equiv\sigma_i^A\otimes I^{B}\otimes I^{C}$, one obtains $W_{\sigma_{i}^{A}}(\mathbf{n}^A,\mathbf{n}^B,\mathbf{n}^C)=\sqrt{3}\,n_{i}^A$. Similarly, if we denote by $\sigma_i^A\sigma_j^B\equiv \sigma_i^A\otimes\sigma_j^B\otimes I^{C}$, then $W_{\sigma_i^A\sigma_j^B}(\mathbf{n}^A,\mathbf{n}^B,\mathbf{n}^C)= 3\,n_{i}^A n_{j}^B$, whereas a three-body operator represented as $\sigma_i^A\sigma_j^B\sigma_k^C\equiv\sigma_{i}^{A}\otimes
\sigma_j^B\otimes\sigma_k^C$ implies that $W_{\sigma_i^A
\sigma_j^B\sigma_k^C}(\mathbf{n}^A,\mathbf{n}^B,\mathbf{n}^C)=3\sqrt{3}\,n_{i}^An_{j}^B n_{k}^C$. Thus, the dependence of the SW symbol on one, two, or three phase space variables distinguishes local contributions from bipartite and tripartite correlations, respectively. This structure will be particularly useful when applying the phase space representation to quantum teleportation. As will be shown below, the information initially encoded in the Wigner function of the input qubit is geometrically redistributed by the unitary operations of the protocol among different correlation sectors of the composite system.

The SW correspondence (\ref{eq:SW3q}) also induces a star product on the product phase space
$\Gamma^{ABC}$. For two arbitrary operators $\hat{M}^{ABC},\hat{N}^{ABC}\in \mathrm{End}(\mathcal{H}^{ABC})$, the star product of their corresponding SW symbols is defined by
\begin{equation}
    W_{\hat{M}^{ABC}}\star W_{\hat{N}^{ABC}}
    :=W_{\hat{M}^{ABC}\hat{N}^{ABC}}.
    \label{eq:tqstar}
\end{equation}
Thus, the SW correspondence maps the noncommutative operator algebra on $\mathcal{H}^{ABC}$ onto a noncommutative algebra of functions on the three qubit phase space. Given the tensor product structure of the SW kernel in Eq.~(\ref{eq:three_qubit_kernel}), consider two factorized operators on $\mathcal{H}^{ABC}$,
\begin{equation}
    \hat{M}^{ABC}
    =\hat{M}^A\otimes\hat{M}^B\otimes\hat{M}^C,
    \qquad
    \hat{N}^{ABC}=\hat{N}^A\otimes\hat{N}^B\otimes\hat{N}^C,
\end{equation}
their SW symbols factorize accordingly,
\begin{equation}
    W_{\hat{M}^{ABC}}=W_{\hat{M}^A}(\mathbf n^A)
    W_{\hat{M}^B}(\mathbf n^B)W_{\hat{M}^C}(\mathbf n^C),
\end{equation}
and similarly for the operator $\hat{N}^{ABC}$. Since $\hat{M}^{ABC}\hat{N}^{ABC}
    =(\hat{M}^A\hat{N}^A)\otimes(\hat{M}^B\hat{N}^B)\otimes(\hat{M}^C\hat{N}^C)$, the corresponding three qubit star product factorizes as \\
\begin{eqnarray} \label{3qSP}
    W_{\hat{M}^{ABC}}
    \star
    W_{\hat{N}^{ABC}}(\mathbf{n}^{A},\mathbf{n}^{B},\mathbf{n}^{C})
    &=&
    \left(
        W_{\hat{M}^A}\star^A W_{\hat{N}^A}
    \right)(\mathbf n^A)
    \\
    &&\times
    \left(
        W_{\hat{M}^B}\star^B W_{\hat{N}^B}
    \right)(\mathbf n^B)
    \left(
        W_{\hat{M}^C}\star^C W_{\hat{N}^C}
    \right)(\mathbf n^C),\nonumber
   \end{eqnarray}
where $\star^j$ denotes the single qubit star product on the coadjoint orbit $(S^2)^j$ as obtained in Eq.~(\ref{starQ}).  By bilinearity, this construction extends naturally to arbitrary three-qubit SW symbols.

The star algebra on the product phase space $\Gamma^{ABC}$ provides an exact phase space realization of the full operator algebra of the three qubit system while preserving its subsystem structure. In particular, SW symbols associated with operators acting on distinct qubits commute under the star product, whereas the noncommutative structure of each individual qubit is entirely captured by its corresponding local star product. This factorized structure allows multipartite unitary transformations to be represented directly in terms of star products of SW symbols on $\Gamma^{ABC}$.

\section{Quantum teleportation in the phase space representation}
\label{sec:teleportation_PS}

With the phase space representation for a general three-qubit system at hand, we now apply this framework to the quantum teleportation protocol. We denote the input qubit by $A$, while $B$ and $C$ constitute the entangled resource shared by Alice and Bob, respectively. The purpose of this section is to formulate the teleportation protocol directly in terms of functions on the product phase space $\Gamma^{ABC}$ and the associated star product, and to analyze how the phase space information carried by the input state is geometrically redistributed through the unitary, measurement, and correction stages of the protocol.

Let the state to be teleported be an arbitrary qubit state with density matrix
\begin{equation}
    \hat{\rho}^A=\frac{1}{2}\left(I^{A}+\mathbf{r}\cdot\boldsymbol{\sigma}^{A}\right),
    \qquad
    \mathbf{r}=(r_x,r_y,r_z),
    \label{eq:input_state}
\end{equation}
where $|\mathbf{r}|\leq 1$. Then, its Wigner function on the coadjoint orbit $(S^2)^A$ is given by
\begin{equation}
    W_{\hat{\rho}^{A}}(\mathbf{n}^A)=\frac{1}{2}\left[1+\sqrt{3}\,\mathbf{r}\cdot\mathbf{n}^A\right].
    \label{eq:input_WF}
\end{equation}
This means that the information specifying the input state is initially represented locally through the dependence of $W_{\hat{\rho}^{A}}(\mathbf{n}^A)$ on the phase space coordinates $\mathbf{n}^A$. The entangled resource shared by qubits $B$ and $C$ is chosen to be the Bell state
\begin{equation}\label{Bell_resource}
    \ket{\Phi^+}^{BC}=\frac{1}{\sqrt{2}}\left(\ket{00}+\ket{11}\right),
\end{equation}
whose density operator admits the following Pauli decomposition
\begin{equation}
    \hat{\rho}_{\Phi^+}^{BC}=\frac{1}{4}\left(I^BI^C+\sigma_{x}^B\sigma_{x}^C-\sigma_{y}^B\sigma_{y}^C
    +\sigma_{z}^B\sigma_{z}^C
    \right).
    \label{eq:Bell_density}
\end{equation}
Applying the SW correspondence (\ref{eq:SW3q}), we obtain
\begin{equation}
    W_{\hat{\rho}_{\Phi^+}^{BC}}(\mathbf{n}^B,\mathbf{n}^C)= \frac{1}{4}
    \left[1+3(\mathbf{n}^{B})^{T}T_{\Phi^+}\mathbf{n}^C\right],
    \label{eq:Bell_WF}
\end{equation}
where $T_{\Phi^+}=\operatorname{diag}(1,-1,1)$, is the correlation matrix associated to the Bell state $\ket{\Phi^+}$. The joint dependence of $W_{\hat{\rho}_{\Phi^+}^{BC}}(\mathbf{n}^B,\mathbf{n}^C)$ on the phase space variables $\mathbf{n}^B$ and $\mathbf{n}^C$ reflects the nonlocal correlations between the two subsystems that constitute the entangled resource.
Since the input state is initially uncorrelated with the Bell pair, the complete density operator takes the form $\hat{\rho}_0=\hat{\rho}^A\otimes\hat{\rho}_{\Phi^+}^{BC}$, and the tensor product property of the SW correspondence (\ref{eq:SW3q}) yields
\begin{equation}
    W_{\hat{\rho}_{0}}(\mathbf{n}^{A},\mathbf{n}^B,\mathbf{n}^C)
    =W_{\hat{\rho}_{A}}(\mathbf{n}^{A})
    W_{\hat{\rho}_{\Phi^+}^{BC}}(\mathbf{n}^B,\mathbf{n}^C).
    \label{eq:initial_SW}
\end{equation}
At this stage, the information carried by the Bloch vector $\mathbf{r}$ is confined to the local function associated with $(S^2)^A$, while the correlations between subsystems $B$ and $C$ are contained in the phase space representation of the Bell-state resource.

The teleportation protocol then redistributes this initially local information across the composite phase space, beginning with a controlled-NOT operation from $A$ to $B$, followed by a Hadamard transformation on qubit $A$. The controlled-NOT operator can be written as the unitary operator
\begin{eqnarray}
\hat{U}_{\mathrm{CNOT}}&=&\ket{0}\bra{0}^A\otimes I^B+
    \ket{1}\bra{1}^A\otimes\sigma_{x}^{B}\\
    &=&
    \frac{1}{2}\left(I^AI^B+\sigma_{z}^{A}I^B+I^A\sigma_{x}^{B}-\sigma_{z}^{A}\sigma_{x}^{B}
    \right),
\label{eq:CNOT}
\end{eqnarray}
while the Hadamard operator reads $\hat{H}^A=\frac{1}{\sqrt{2}}(\sigma_{x}^A+\sigma_{z}^A)$.
The three-qubit state immediately before Alice's measurement is therefore
\begin{equation}
    \hat{\rho}_1=(\hat{H}^A\otimes I^B\otimes I^C)\hat{U}_{\mathrm{CNOT}}\hat{\rho}_0 \hat{U}_{\mathrm{CNOT}}^\dagger
    (\hat{H}^A\otimes I^B\otimes I^C).
    \label{rho1}
\end{equation}
Under the SW correspondence (\ref{eq:SW3q}), the operator multiplication is represented by the star product (\ref{3qSP}), hence, the phase space representation of $\hat{\rho_{1}}$ becomes
\begin{equation}
    W_{ \hat{\rho}_1}(\mathbf{n}^{A},\mathbf{n}^B,\mathbf{n}^C)=W_{\hat{H}^A}
    \star W_{\hat{U}_{\mathrm{CNOT}}}
    \star W_{\hat{\rho}_{0}}
    \star W_{\hat{U}_{\mathrm{CNOT}}}^*
    \star W_{\hat{H}^A},
    \label{W1_star}
\end{equation}
where the phase space symbols associated with $\hat{U}_{\mathrm{CNOT}}$ and $\hat{H}^{A}$ are given, respectively, by
\begin{equation}
W_{\hat U_{\mathrm{CNOT}}}
(\mathbf n^A,\mathbf n^B)=\frac{1}{2}\Big[1+\sqrt{3}n_z^A+\sqrt{3}n_x^B-3n_z^A n_x^B
\Big],
\label{SW_NOT}
\end{equation}
and
\begin{equation}
W_{\hat H_A}(\mathbf n^A)=\sqrt{\frac{3}{2}}\left(n_x^A+n_z^A\right).
\label{SW_H}
\end{equation}

Substituting these expressions into Eq.~(\ref{W1_star}) and evaluating the star products
according to Eq.~(\ref{3qSP}), we obtain

\begin{align}
 W_{ \hat{\rho}_1}(\mathbf{n}^{A},\mathbf{n}^B,\mathbf{n}^C)=\frac18\Big\{&
1+3n_{x}^{B}n_{x}^{C}
-3r_z n_{y}^{B}n_{y}^{C}
+3r_z n_{z}^{B}n_{z}^{C} \nonumber\\
&+\sqrt3\,r_z n_{x}^{A}
+3\sqrt3\,r_z n_{x}^{A}n_{x}^{B}n_{x}^{C} \nonumber\\
&-3\sqrt3\,n_{x}^{A}n_{y}^{B}n_{y}^{C}
+3\sqrt3\,n_{x}^{A}n_{z}^{B}n_{z}^{C} \nonumber\\
&-3r_y n_{y}^{A}n_{x}^{C}
-3r_y n_{y}^{A}n_{x}^{B} \nonumber\\
&+3\sqrt3\,r_x n_{y}^{A}n_{y}^{B}n_{z}^{C}
+3\sqrt3\,r_x n_{y}^{A}n_{z}^{B}n_{y}^{C} \nonumber\\
&+3r_x n_{z}^{A}n_{x}^{C}
+3r_x n_{z}^{A}n_{x}^{B} \nonumber\\
&+3\sqrt3\,r_y n_{z}^{A}n_{y}^{B}n_{z}^{C}
+3\sqrt3\,r_y n_{z}^{A}n_{z}^{B}n_{y}^{C}
\Big\}.
\label{Wrho1}
\end{align}
The preceding expression provides the complete phase space representation of the three qubit state immediately before measurement. In particular, the components of the input Bloch vector $\mathbf{r}=(r_x,r_y,r_z)$ now appear as coefficients of functions involving phase space coordinates associated with different coadjoint orbits. For instance, the term $r_x n_{z}^{A}n_{x}^{C}$ contributes to the bipartite correlation sector involving qubits $A$ and $C$, whereas
$r_x n_{y}^{A}n_{y}^{B}n_{z}^{C}$ contributes to the tripartite correlation sector. Thus, the phase space representation explicitly reveals how the information initially confined to the local phase space $(S^2)^A$ is redistributed among the correlations of the composite three qubit system.

It is important to emphasize that this redistribution does not imply that the input state has already become locally available to Bob. Taking the partial trace over Alice's qubits gives
\begin{equation}
    \hat{\rho}_{1}^{C}=
    \tr^{AB}\hat{\rho}_{1}=\frac{1}{2}I^{C},
    \label{eq:Bob_before_measurement}
\end{equation}
where the partial trace, $\tr^{AB}:\mathrm{End}(\mathcal{H}^{A}\otimes \mathcal{H}^{B}\otimes \mathcal{H}^{C})\to \mathrm{End}(\mathcal{H}^{C})$, is taken over the Hilbert spaces of qubits $A$ and $B$, leaving an operator acting on $\mathcal{H}^{C}$. Therefore, its corresponding Wigner function takes the form
\begin{equation}
    W_{ \hat{\rho}_{1}^{C}}(\mathbf{n}_C)
    = \int_{(S^2)^{A}\times (S^{2})^{B}}W_{ \hat{\rho}_1}(\mathbf{n}^{A},\mathbf{n}^B,\mathbf{n}^C)d\mu^{A}d\mu^{B}=
    \frac{1}{2}.
    \label{eq:Bob_SW_before_measurement}
\end{equation}
Consequently, prior to Alice's measurement, Bob's local Wigner function is independent of the input Bloch vector $\mathbf r$. The information characterizing the input state is therefore present exclusively in the correlation structure of the full three qubit Wigner function $W_{ \hat{\rho}_1}(\mathbf{n}^{A},\mathbf{n}^B,\mathbf{n}^C)$.

Following the unitary transformations, Alice performs a projective
measurement on qubits $A$ and $B$ in the computational basis. Each measurement outcome $(a,b)$, with $a,b\in\{0,1\}$, is associated with the $\hat{P}_{ab}=\ket{ab}\bra{ab}$.
The resulting unnormalized conditional state of qubit $C$ is
obtained by projecting the three qubit density operator $\hat{\rho}_1$ onto the measurement outcome and subsequently
taking the partial trace over Alice's subsystems,
\begin{equation}
    \hat{\rho}^C_{(ab)}=\tr^{AB}
    \left[
        (\hat{P}_{ab}\otimes I^C)
        \hat{\rho}_1
        (\hat{P}_{ab}\otimes I^C)
    \right].
    \label{conditional_density}
\end{equation}

The SW correspondence provides a phase space representation of this measurement procedure.
Since, the operator multiplication is represented by the star product, whereas the partial trace over $A$ and $B$ corresponds to integration over their respective coadjoint orbits. Therefore, the unnormalized conditional Wigner function of qubit $C$, denoted by $\widetilde{W}_{\hat{\rho}^C_{(ab)}}(\mathbf{n}^C)$, takes the form
\begin{equation}
\widetilde{W}_{\hat{\rho}^C_{(ab)}}(\mathbf{n}^C)
    =
    \int_{(S^2)^A\times (S^2)^B}
    \left(
        W_{\hat{P}_{ab}}
        \star W_{\hat{\rho_{1}}}
        \star W_{\hat{P}_{ab}}
    \right)d\mu^A
     d\mu^B.
    \label{eq:projection_SW}
\end{equation}
Here, $W_{\hat{P}_{ab}}$ denotes the SW symbol of the measurement projector $\hat{P}_{ab}$, given by
\begin{equation}
    W_{\hat{P}_{ab}}(\mathbf{n}^A,\mathbf{n}^B)=
    \frac{1}{4}\left[1+(-1)^a\sqrt{3}n_{z}^{A}\right]\left[
        1+(-1)^b\sqrt{3}n_{z}^{B}\right].
    \label{measurement_projector_SW}
\end{equation}
The probability associated with each measurement outcome is determined by the trace of the corresponding unnormalized conditional density operator. By the standardization property of the SW correspondence, this probability can equivalently be expressed as
\begin{equation}
    p_{ab}=
    \tr\,\hat\rho^C_{(ab)}=
    \int_{(S^2)^C} \widetilde{W}_{ \hat\rho^C_{(ab)}}(\mathbf{n}^C)d\mu^C.
    \label{eq:measurement_probability_SW}
\end{equation}
For the maximally entangled resource state considered in Eq.~(\ref{Bell_resource}), all four outcomes occur with equal probability, $p_{ab}=1/4$.
The corresponding normalized conditional Wigner function of the qubit $C$ appearing in~(\ref{eq:measurement_probability_SW}) is therefore given by $W_{\hat{\rho}^C_{(ab)}}(\mathbf{n}^C)=\widetilde{W}_{\hat{\rho}^C_{(ab)}}(\mathbf{n}^C)/p_{ab}$.

Evaluating Eq.~(\ref{eq:projection_SW}) for each of the four
possible measurement outcomes, we obtain the corresponding normalized Wigner functions, given by
\begin{equation}
    W_{\hat{\rho}^C_{(ab)}}(\mathbf{n}^C)=\frac{1}{2}\left[1+\sqrt{3}\left((-1)^{a}r_{x}n^{C}_{x}+(-1)^{a+b}r_{y}n^{C}_{y}+(-1)^{b}r_{z}n_{z}^{C}\right)\right].
\end{equation}
These expressions show that each measurement outcome
yields a conditional Wigner function that preserves the dependence on the input Bloch vector, with possible sign changes in its components. 
The geometrical interpretation of these conditional
states follows from the covariance property of the
SW correspondence. In particular, introducing the rotations
\begin{equation}
    R_{ab}=\left(R_{x}(\pi)\right)^{b}\left(R_{z}(\pi)\right)^{a}, \qquad a,b\in \{0,1\},
\end{equation}
the four components of the conditional Wigner function can be expressed
in the unified form
\begin{equation}
     W_{\hat{\rho}^C_{(ab)}}(\mathbf{n}^C)
    =
    W_{\hat{\rho}^{A}}(R_{ab}^{-1}\mathbf{n}^C).
    \label{eq:conditional_rotation}
\end{equation}
Thus, each measurement outcome determines a rotation of the initial Wigner function, yielding a conditional phase space configuration on Bob's coadjoint orbit that preserves the complete information encoded in the input state. Although Bob's reduced Wigner function is independent of the input Bloch vector prior to measurement, the conditional Wigner functions recover its full dependence on $\mathbf{r}$. In this way, Alice's measurement extracts a local phase space configuration from the multipartite correlation structure of the three qubit system, while the corresponding measurement outcome identifies the rotation to be undone by Bob's subsequent Pauli correction.

Finally, upon receiving Alice's classical measurement outcome $(a,b)$, Bob applies the corresponding Pauli correction
\begin{equation}
    \hat U_{ab}=(\sigma_{x}^{C})^b(\sigma_{z}^{C})^a,
    \qquad a,b\in\{0,1\},
    \label{eq:Pauli_corrections}
\end{equation}
to his conditional state $\hat{\rho}^{C}_{(ab)}$  thereby reversing the outcome-dependent rotation and recovering the initial quantum state. Under the SW correspondence, this unitary transformation is represented by the star-adjoint action
\begin{equation}
\begin{split}
    W_{\hat{\rho}^{C}_{\mathrm{out}}}(\mathbf n^C)
    &=
    W_{\hat U_{ab}}
    \star W_{\hat{\rho}^{C}_{(ab)}}
    \star W_{\hat U_{ab}^{\dagger}}
    \\
    &=
    W_{\hat{\rho}^{C}_{(ab)}}
    (R_{ab}^{-1}\mathbf n^C),
\end{split}
    \label{eq:Pauli_correction_SW}
\end{equation}
where the second equality follows from the covariance property of the SW correspondence (\ref{eq:SW_rotation_covariance}). Substituting Eq.~(\ref{eq:conditional_rotation}) into Eq.~(\ref{eq:Pauli_correction_SW}), we obtain
\begin{equation}
\begin{split}
    W_{\hat{\rho}^{C}_{\mathrm{out}}}(\mathbf n^C)
    &=
    W_{\hat{\rho}^{A}}
    (R_{ab}^{-1}R_{ab}^{-1}\mathbf n^C)
    \\
    &=
    W_{\hat{\rho}^{A}}(\mathbf n^C),
\end{split}
    \label{eq:teleportation_identity}
\end{equation}
where we have used $R_{ab}^{-1}=R_{ab}$ and $R_{ab}^{2}=I$.

Thus, independently of Alice's measurement outcome, Bob's final Wigner function coincides with that of the initial state. By the invertibility of the SW correspondence (\ref{eq:SW_inverse}), this result is equivalent to $\hat{\rho}^{C}_{\mathrm{out}}=\hat{\rho}^{A}$, confirming the successful teleportation of the input quantum state. In the phase space formulation, the teleportation protocol provides a geometrical description of how the information initially represented in the Wigner function of qubit $A$ is redistributed among the multipartite correlations of the three qubit system. Following Alice's measurement, this information is recovered in the form of an outcome-dependent Wigner function on Bob's coadjoint orbit, whose original orientation is restored through the appropriate Pauli correction.


\section{Geometrical interpretation and extensions to other phase spaces geometries}
\label{sec:geometrical_interpretation}

The phase space formulation developed in the preceding section provides a geometrical interpretation of quantum teleportation in terms of the symplectic structure of coadjoint orbits and the noncommutative algebra of their associated SW symbols. In contrast to the conventional Hilbert space description, where the protocol is formulated through operations on density operators, the present approach represents the evolution of quantum information through functions defined on the product of symplectic manifolds. This allows the different stages of teleportation to be understood in terms of the geometric redistribution of phase space information, its conditional recovery, and its subsequent reconstruction on the receiver's coadjoint orbit.

As illustrated by the teleportation protocol, the phase space description of a single qubit is naturally formulated on the coadjoint orbit $S^2$ of $SU(2)$, endowed with KKS symplectic form $\omega_{\mathrm{KKS}}$ defined in Eq.~(\ref{KKS}). Within this framework, the corresponding Wigner function is not an arbitrary function on the sphere, but belongs to the finite dimensional space of SW symbols associated with operators acting on $\mathbb C^2$. In particular, as shown in Eq.~ (\ref{eq:qubit_Wigner}), the Wigner function completely characterizes the quantum state through its Bloch vector $\mathbf{r}$.  Thus, the symplectic manifold provides the geometric setting for the phase space representation of quantum states, while the star product reproduces the noncommutative algebra of quantum operators.

For the three qubit system, the relevant phase space is the product manifold $\Gamma^{ABC}$ defined in Eq.~(\ref{eq:three_qubit_phase_space}), equipped with the product symplectic structure
\begin{equation}
    \omega^{ABC}
    =
    \pi_{A}^{*}\omega^{A}_{\mathrm{KKS}}
    +
    \pi_{B}^{*}\omega^{B}_{\mathrm{KKS}}
    +
    \pi_{C}^{*}\omega^{C}_{\mathrm{KKS}},
    \label{eq:product_symplectic_form}
\end{equation}
where $\pi_{j}:\Gamma^{ABC}\to(S^2)^j$ denotes the canonical projection onto the $j$-th factor, and $\pi_{j}^*$ is the corresponding pullback of differential forms. The resulting symplectic structure is the direct sum of the KKS forms associated with the individual qubits. In particular, it preserves the independent symplectic geometry of each subsystem, while providing a common phase space on which multipartite quantum correlations can be represented through the tensor product and its associated star product.

An important feature of the teleportation protocol is the redistribution of quantum information among the different coadjoint orbits. Initially, the input state is encoded in the local Wigner function $W_{\hat\rho^A}(\mathbf n^A)$. After the controlled-NOT and Hadamard transformations, this information is encoded in the multipartite correlations of the three qubit system, while Bob's reduced Wigner function remains independent of the input Bloch vector. Thus, the phase space formulation provides a geometrical description of how the information initially represented locally becomes distributed among the correlations of the composite system.

Alice's measurement subsequently selects an outcome-dependent conditional Wigner function on Bob's coadjoint orbit. As established in Eq.~(\ref{eq:conditional_rotation}), this function is related to the initial Wigner function by a rotation.
The rotations $R_{ab}\in SO(3)$ associated with the conditional states arise from the coadjoint action of $SU(2)$ on the spherical orbit. Since the KKS symplectic form is invariant under this action, these rotations are symplectomorphisms satisfying
\begin{equation}
    R_{ab}^{*}\omega_{\mathrm{KKS}}
    =
    \omega_{\mathrm{KKS}}.
    \label{eq:rotation_symplectomorphism}
\end{equation}
Thus, the conditional Wigner functions are related by transformations that preserve the underlying symplectic geometry of the coadjoint orbit. The conditional states therefore correspond to different orientations of the same  phase space function, related through transformations that preserve the underlying symplectic structure. The Pauli correction reverses the corresponding rotation, reconstructing the original Wigner function on his coadjoint orbit.

This interpretation requires a distinction between the underlying geometry of the phase space and the quantum information represented on it. The teleportation protocol does not transfer the coadjoint orbit, its KKS symplectic form, or the associated star product algebra from Alice to Bob. These structures are already present in the phase space descriptions of both subsystems. Rather, the protocol reconstructs the Wigner function on Bob's orbit, preserving the information encoded in the original quantum state. In this sense, the geometrical content of teleportation lies in the redistribution and reconstruction of quantum state information within a common symplectic and noncommutative framework.

The preceding construction also suggests a more general formulation of quantum teleportation within the framework of deformation quantization. Consider a quantum system associated with a symplectic manifold $(\mathcal M,\omega)$ and a suitable invertible symbol correspondence $\mathcal W:
\hat A\longmapsto W_{\hat A}$, whose image carries an associative star product satisfying $W_{\hat A\hat B}=W_{\hat A}\star W_{\hat B}$ \cite{Bayen,Fedosov,Kontsevich}.
For three equivalent subsystems, the corresponding phase space description is constructed on $\mathcal M^A\times\mathcal M^B\times\mathcal M^C$, equipped with the product symplectic structure. When the symbol correspondence is compatible with tensor products and partial traces, the unitary and measurement stages of a teleportation protocol can be expressed through star products and integrations over the appropriate phase space variables.

In particular, consider a measurement outcome $\alpha$ associated with a projector $\hat P_\alpha$ acting on Alice's subsystems. Following Eq.~(\ref{eq:projection_SW}), the corresponding conditional Wigner function on Bob's phase space is obtained by integrating over the phase space variables of the measured subsystems,
\begin{equation}
W_{\hat{\rho}^{C}_{\alpha}}(x^C)
    =
    \frac{1}{p_\alpha}
    \int_{\mathcal M^A\times\mathcal M^B}
    \left[
        W_{\hat P_\alpha}
        \star W_{\hat{\rho}_{1}}
        \star W_{\hat P_\alpha}
    \right]
    d\mu_A\,d\mu_B,
    \label{eq:general_conditional_symbol}
\end{equation}
where $x^C$ stands for the local coordinates of the phase space $\Gamma^C$ for the 
subsystem $C$, and $p_\alpha$ denotes the probability of obtaining the measurement outcome $\alpha$. The subsequent correction is implemented through the star-adjoint action of the corresponding unitary operator $\hat U_\alpha$,
\begin{equation}
    W_{\hat\rho^C_{\mathrm{out}}}
    =
    W_{\hat U_\alpha}
    \star W_{\hat\rho^C_\alpha}
    \star W_{\hat U_\alpha^\dagger}.
    \label{eq:general_star_correction}
\end{equation}
When $\hat U_\alpha$ belongs to a symmetry group whose action is represented covariantly on the phase space, the correction admits a geometric interpretation as the pullback of the conditional Wigner function under the corresponding symplectomorphism. The rotations of the coadjoint orbit $S^2$ arising in the qubit teleportation protocol provide a particularly simple realization of this general mechanism.

A natural extension of this framework involves higher-spin systems, whose phase spaces are also described by coadjoint orbits of $SU(2)$. Although the underlying manifold remains a two-dimensional sphere, both the SW correspondence and the associated star product depend on the spin representation \cite{Rundle,Klimov}. In contrast to the qubit case, where the SW symbols involve only constant and linear terms, higher-spin representations incorporate spherical harmonics of higher orders, reflecting the additional multipole structure of the quantum states. This provides a promising framework for investigating how higher-order multipole information is redistributed through multipartite phase space correlations during quantum teleportation.

Another possible extension concerns finite-dimensional quantum systems associated with higher symmetry groups. For a $d$-level system, the coadjoint orbits of $SU(d)$ provide natural symplectic phase spaces endowed with the KKS symplectic form. Depending on the choice of orbit, these manifolds can be identified with complex projective spaces or, more generally, flag manifolds~\cite{Flag}. Within this geometric setting, an appropriate SW correspondence would allow the formulation of quantum teleportation in terms of phase space functions defined on these higher-dimensional manifolds, while the conditional unitary corrections could be interpreted through the induced action of the corresponding symmetry group.

A complementary extension involves continuous-variable quantum systems, whose canonical phase space $\mathbb{R}^{2n}$ is endowed with the standard symplectic form, and whose operator algebra is represented through the Moyal star product \cite{Bayen,Zachos,Dito}. In this setting, phase space translations and linear symplectic transformations play a role analogous to that of rotations on the qubit coadjoint orbit. Continuous variable quantum teleportation thus provides a natural framework for investigating how quantum information is transferred through conditional phase space transformations and subsequently recovered by appropriate corrections, within the broader setting of deformation quantization.

These examples suggest that the phase space formulation of quantum teleportation is not limited to the spherical geometry of qubits. However, the existence of a symplectic or Poisson manifold equipped with an associative star product is not sufficient to ensure a deterministic teleportation protocol. Its implementation also requires an appropriate entangled resource, a suitable measurement scheme, and conditional operations that allow the reconstruction of the unknown input state. Consequently, extending this framework to other symplectic and Poisson geometries requires considering not only the mathematical properties of the corresponding symbol calculus, but also the quantum information resources and operations available in each system.

\section{Conclusions}
\label{sec:conclusions}

In this work, we have formulated the quantum teleportation protocol within the framework of deformation quantization, using the Stratonovich--Weyl correspondence and the associated star product on the coadjoint orbits of $SU(2)$. By representing the three qubit system on a phase space product, we have shown how the information initially encoded in the Wigner function of the input qubit is redistributed among multipartite correlations, recovered through conditional measurements, and reconstructed on Bob's coadjoint orbit by applying the appropriate Pauli correction.

The geometrical interpretation of the protocol follows from the covariance of the SW correspondence, which relates the conditional Wigner functions through rotations that preserve the KKS symplectic structure. From this perspective, quantum teleportation can be understood as the reconstruction of the input state's Wigner function on the receiver's coadjoint orbit, rather than the transfer of the underlying symplectic geometry itself.

From this perspective, the present construction provides a starting point for investigating quantum teleportation on different phase space geometries, including higher-spin systems, higher-dimensional coadjoint orbits, and continuous-variable systems. A natural direction for future research is to investigate how the reconstruction of quantum information in these settings depends on the underlying symplectic geometry, the associated star product, and the representation of the relevant symmetry group. Various formal developments in this direction will be reported in a future communication.

\section*{Acknowledgments}
The authors would like to acknowledge support from SNII SECIHTI-Mexico. JBM acknowledge financial support from Marcos Moshinsky foundation. AM acknowledges financial support from COPOCYT under project 2467 HCDC/2024/SE-02/16 (Convocatoria 2024-03, Fideicomiso 23871).

\section*{Data availability statement}

No Data associated in the manuscript


\bibliographystyle{unsrt}

\end{document}